\documentclass[fleqn,usenatbib]{mnras}

\usepackage{newtxtext,newtxmath}

\usepackage[T1]{fontenc}

\DeclareRobustCommand{\VAN}[3]{#2}
\let\VANthebibliography\thebibliography
\def\thebibliography{\DeclareRobustCommand{\VAN}[3]{##3}\VANthebibliography}

\newcommand{\src}{V4641~Sgr}

\newcommand{\xmm}{\textit{XMM-Newton}}
\newcommand{\chandra}{\textit{Chandra}}
\newcommand{\xrism}{\textit{XRISM}}

\usepackage{graphicx}	
\usepackage{amsmath}	

\title[RGS spectrum of \src]{Dense, multi-phase accretion disk atmosphere in the low-luminosity state of black hole transient \src}

\author[Zhang et al.]{Zuobin Zhang,$^{1}$\thanks{E-mail: zuobin.zhang@physics.ox.ac.uk}
Rob Fender,$^{1,2}$
James H. Matthews,$^{1}$
Jiachen Jiang,$^{3}$
Honghui Liu,$^{4}$  \newauthor
Alessandra Ambrifi,$^{5,6}$ 
Teo Mu\~{n}oz-Darias,$^{5,6}$
Maxime Parra,$^{7}$
Megumi Shidatsu,$^{7}$
Menglei Zhou,$^{4}$ \newauthor
Yuexin Zhang,$^{8,9}$ 
Abdurakhmon~Nosirov,$^{10}$
Cosimo Bambi,$^{10,11}$ 
and Justine Crook-Mansour,$^{1}$
\\
$^{1}$Astrophysics, Department of Physics, University of Oxford, Keble Road, Oxford OX1 3RH, UK\\
$^{2}$Department of Astronomy, University of Cape Town, Private Bag X3, Rondebosch 7701, South Africa\\
$^{3}$Department of Physics, University of Warwick, Gibbet Hill Road, Coventry CV4 7AL, UK\\
$^{4}$Institut f\"ur Astronomie und Astrophysik, Eberhard-Karls Universit\"at T\"ubingen, D-72076 T\"ubingen, Germany\\
$^{5}$Instituto de Astrof\'{i}sica de Canarias, E-38205 La Laguna, Tenerife, Spain\\
$^{6}$Departamento de astrof\'{i}sica, Universidad de La Laguna, E-38206 La Laguna, Tenerife, Spain\\
$^{7}$Department of Physics, Ehime University, 2-5, Bunkyocho, Matsuyama, Ehime 790-8577, Japan\\
$^{8}$Center for Astrophysics, Harvard \& Smithsonian, 60 Garden St, Cambridge, MA 02138, USA\\
$^{9}$Kapteyn Astronomical Institute, University of Groningen, P.O.\ BOX 800, 9700 AV Groningen, The Netherlands\\
$^{10}$Center for Astronomy and Astrophysics, Center for Field Theory and Particle Physics, Fudan University, Shanghai 200438, China\\
$^{11}$School of Natural Sciences and Humanities, New Uzbekistan University, Tashkent 100000, Uzbekistan
}

\date{Accepted XXX. Received YYY; in original form ZZZ}

\pubyear{\the\year{}}

\begin{document}
\label{firstpage}
\pagerange{\pageref{firstpage}--\pageref{lastpage}}
\maketitle

\begin{abstract}
We present soft X-ray spectroscopy of the black-hole X-ray binary \src\ with the \xmm\ Reflection Grating Spectrometer (RGS). The RGS spectrum shows narrow emission features from N\,\textsc{vi--vii} and O\,\textsc{vii--viii} superimposed on a partially covered disk blackbody continuum. A blind Gaussian search confirms the presence of significant lines at the expected rest wavelengths. He-like triplet ratios (high $G$, low $R$) and full photoionization modelling both indicate a dense, photoionized plasma. Small redshifted velocities of $\sim 540$--$720\ \mathrm{km\ s^{-1}}$ are suggested, which are consistent with quasi-static or slowly flowing gas away from the observer after accounting for systematics. Photoionization modelling requires two \textsc{xstar} components with an intermediate ionization parameter ($\log\xi \simeq 3.1$) and a low ionization parameter ($\log\xi \simeq 0.36$), respectively. The simultaneous EPIC-pn spectrum suggests highly ionized Fe emission structures, hinting at an additional, more highly ionized component. These results imply the existence of a radially extended, multiphase, and dense disk atmosphere in the source. We compare the source with other X-ray binaries showing similar emission lines. \src\ shares a similarly high inclination with other sources; however, the presence of low ionization emission lines distinguishes it from the rest.

\end{abstract}




\section{Introduction} \label{sec:intro}

X-ray binaries (XRBs) are among the most powerful astrophysical laboratories for studying accretion and outflows around compact objects (e.g. \citealt{Done2007, Kording2006, Zhang2022, Zhang2021}). In these systems, matter is transferred from a stellar companion onto a neutron star or black hole through an accretion disk, producing strong X-ray emission. X-ray observations of XRBs reveal two canonical accretion states----the spectrally hard and soft states \citep{Remillard2006}. The hard state is characterized by a hot, Comptonizing corona that supplies most of the X-ray flux \citep{Sunyaev1979}, whereas in the soft state, the corona component is suppressed and the emission is dominated by a geometrically thin, optically thick accretion disk \citep{Novikov1973, Shakura1973}. 

Intense radiation from the inner disk irradiates the outer regions, resulting in atmospheric layers above the disk, where external radiative heating can exceed local viscous dissipation, reshaping the thermal/ionization balance and vertical structure of the plasma (e.g. \citealt{Diaz2016, Jimenez-Garate2002}). The disk atmosphere is critical for interpreting both the spectral and dynamical properties. For example, deviations from the standard multicolor blackbody model are introduced by the vertical structure, scattering processes, and partial ionization of metals in the atmosphere, which are accounted for by a color correction factor in the disk emission (e.g. \citealt{Shimura1995}). Radiation--MHD simulations further demonstrate that magnetically dominated surface layers and associated instabilities strongly influence radiative transfer \citep{Hirose2006}.

It is commonly believed that disk winds play a significant role in governing accretion dynamics, angular momentum transport, and feedback into the surrounding environment (e.g. \citealt{Zhang2026, Munoz2026, Zhang2024, Ponti2012, Fender2016, Begelman1983}). It remains uncertain whether static disk atmospheres are physically linked to disk winds producing blue-shifted absorptions, or whether they represent distinct phenomena. One possibility is that a static atmosphere can evolve into an outflow under specific physical conditions. Observations of GRS~1915+105 illustrate this diversity: in some epochs, absorption of ionized species is present without measurable blueshift (e.g. \citealt{Lee2002}), whereas in others disk winds are unambiguously detected (e.g. \citealt{Miller2016, Ueda2009, Liu2022}). Exploring the possible connection between disk atmosphere and winds in compact binaries is particularly important to understand the driving mechanisms of disk winds \citep{Diaz2016}.


High-resolution X-ray spectroscopy has revealed that many XRBs display narrow emission or/and absorption lines thought to trace an atmosphere above the disk. These features provide constraints on the ionization state, density, geometry, and kinematics of the disk atmosphere (e.g. \citealt{Diaz2016, Buisson2020, Trueba2020}). In several well-studied neutron star systems, such as EXO~0748--676 and MXB~1659--298, the disk atmosphere has been shown to comprise multiple ionization phases, with densities exceeding $10^{13}$~cm$^{-3}$ and often located within $\sim 10^4-10^5$~$R_{\rm g}$ (where $R_{\rm g} = GM_{\rm BH}/c^2$ is the gravitational radius) from the compact object \citep{Psaradaki2018, Ponti2019}. Observations suggest a sensitive dependence of the atmosphere structure on the spectral energy distribution, possibly attributed to thermal photoionization instabilities \citep{Ponti2019}. Comparable disk atmosphere features have also been found in black hole binaries, such as GRS~1915+105 \citep{Lee2002}. High-resolution spectroscopy has revealed that such an atmosphere can coexist with disk winds, contributing low-velocity line components (e.g. \citealt{Lee2002, Ueda2009, Kallman2009}). These studies demonstrate that disk atmosphere are a recurrent phenomenon in both neutron star and black hole X-ray binaries, and they provide crucial insight into the structure and dynamics of accretion flows.

The Galactic black hole X-ray binary \src\ provides a unique laboratory to investigate these processes. This system, consisting of a $\sim6.4$~$M_{\bigodot}$ black hole accreting from a B9III companion star, is located at a distance of $D = 6.2 \pm 0.7$~kpc \citep{Orosz2001, MacDonald2014}. It exhibits unusual outburst behavior, including extremely rapid and luminous flares as well as soft states at very low Eddington fractions \citep{Revnivtsev2002, Hjellming2000, Bahramian2015}. \src\ has attracted significant attention following the detection of very-high-energy emission by LHAASO \citep{LHAASO2024, Alfaro2024}, making it one of the highest-energy X-ray binaries known to date. This discovery has subsequently motivated searches for extended emission at lower energies, leading to the detection of large-scale structures in the radio \citep{Grollimund2026} and X-ray bands \citep{Suzuki2025}. An optical spectroscopic study suggests the presence of cold wind outflows in the optical band in the low luminosity state of the source \citep{Munoz2018}. The terminal velocity of the cold wind is estimated to lie in the range of $\sim900$–$1600\ \mathrm{km\ s^{-1}}$, with indications that it can reach up to $\sim3000\ \mathrm{km\ s^{-1}}$ in some cases.


\chandra/HETG High-resolution X-ray spectra during the 2020 outburst of \src\ revealed numerous highly ionized metal lines superimposed on a disk-dominated continuum, requiring at least two photoionized plasma components to reproduce the observed spectra \citep{Shaw2022}. More recently, \xrism\ observations during its 2024 outburst have uncovered a complex and variable multiphase flow, including both redshifted and blueshifted components, further emphasizing the inhomogeneous and dynamic nature of the disk atmosphere in this source \citep{Parra2025}. Despite these advances, the physical structure and dynamics of the disk atmosphere in \src\ remain poorly constrained. In particular, it is unclear whether the observed emission lines primarily originate from static atmosphere or the base of an outflowing wind. Addressing these questions is essential for building a unified picture of accretion–outflow coupling in black hole XRBs. In this work, we present a detailed investigation of the disk atmosphere in \src\ with the latest \textit{XMM-Newton}/RGS data. By analyzing line diagnostics and photoionization modelling, we aim to characterize the plasma conditions and probe the physical properties of the disk atmosphere in this unusual system.

The paper is organized as follows. In Sec.~\ref{sec:data}, we present the observational data reduction. The Gaussian line fitting and the photoionization modelling are reported in Sec.~\ref{sec:results}. We discuss the results and report our conclusions in Sec.~\ref{sec:discussion} and Sec.~\ref{sec:conclusion}, respectively.


\section{observation and data reduction} \label{sec:data}

\begin{figure*}
    \centering
    \includegraphics[width=0.9\linewidth]{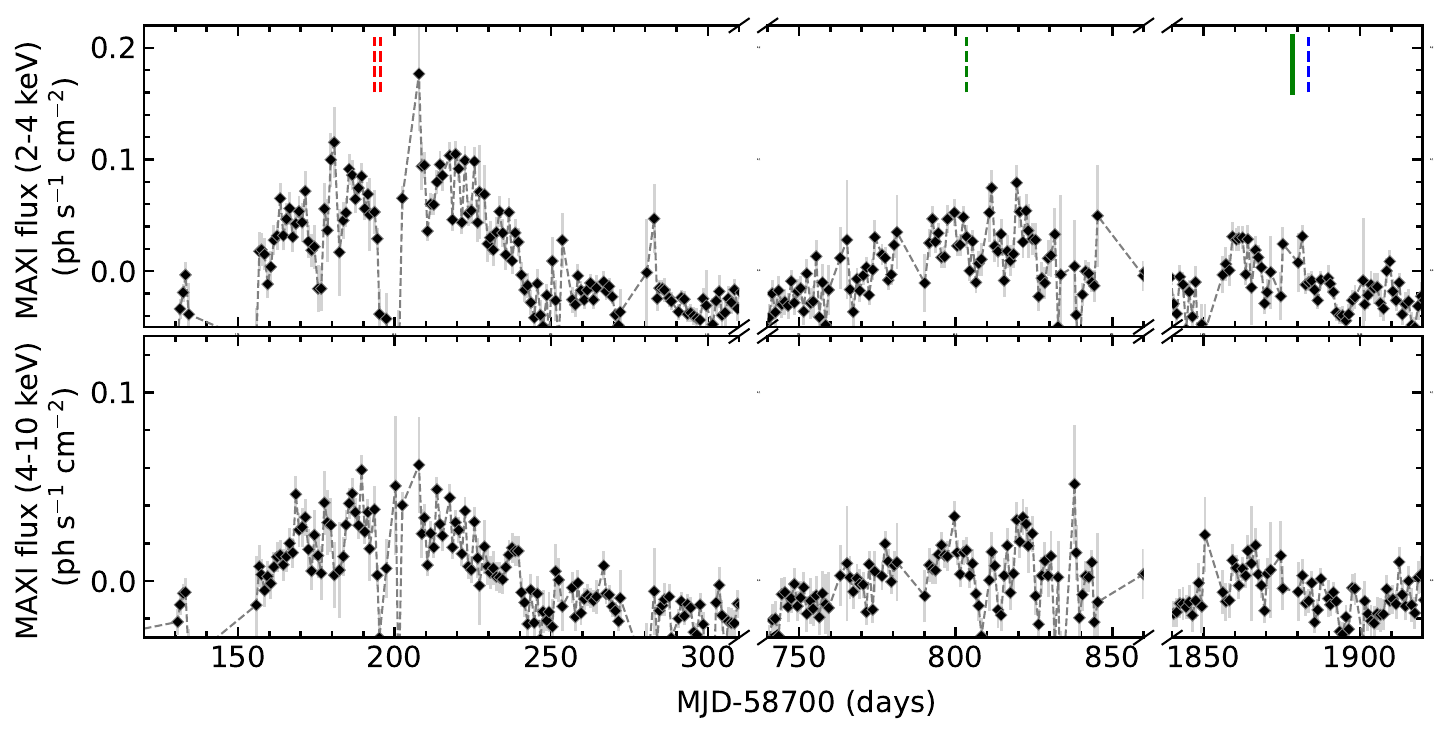} \\
    \caption{MAXI light curves of \src\ in the 2--4~keV (top) and 4--10~keV (bottom) energy bands through the on-demand system \citep{Matsuoka2009}, showing three recent outbursts. Vertical lines indicate the epochs of high-resolution X-ray observations. Green lines mark the XMM-Newton observations, red lines mark the \chandra\ observations, and the blue line marks the \xrism\ observation. This work focuses on the second XMM-Newton dataset, which is emphasized accordingly. The x-axis represents time in units of MJD (Modified Julian Date)- 58700 days.}
    \label{maxi_lc}
\end{figure*}

In response to the bursting activity of \src\ since September 2024 \citep{Negoro2024}, \xmm\ observed the source on September 25, 2024 (Obs ID: 0953011901), as part of a Target of Opportunity (ToO) program. This observation was triggered shortly after the reported outburst onset (Figure~\ref{maxi_lc}) and aimed to capture potential soft X-ray emission/absorption features during its active state. Compared to earlier observations with \chandra\, the \xmm\ data offer improved sensitivity and a higher effective area in the soft X-ray band. In particular, the observation includes data from the Reflection Grating Spectrometer (RGS; \citealt{den_Herder2001}), a high-resolution soft X-ray spectrometer onboard \xmm\ designed to probe emission and absorption features in the 0.35--2.5~keV range. The RGS data provide an opportunity to investigate the ionized environment around \src\ with spectral resolution higher than that of CCD-based instruments. In this work, we present a detailed analysis of the RGS spectrum of \src, focusing on the revealed accretion or outflow signatures. This latest \xmm\ observation serves as the main dataset for the analysis presented in this work.

The \xmm\ data were reduced with the latest version (19.1.0) of the \xmm\ Science Analysis System (SAS), applying the most recent calibration files. The RGS1 and RGS2 data were processed using the standard \texttt{rgsproc} pipeline. The original exposure times were 50.99~ks and 50.97~ks for RGS1 and RGS2, respectively. To identify intervals affected by flaring particle background, we extracted background light curves from both RGS detectors. Time intervals with background count rates exceeding 0.2~counts s$^{-1}$ were classified as flaring particle background and excluded from the analysis. After applying this filtering, the effective exposure times were reduced to 49.08~ks for RGS1 and 49.02~ks for RGS2. The filtered first-order spectra from both detectors were subsequently combined using the \texttt{rgscombine} task to produce a single stacked spectrum, which was adopted for the subsequent analysis.

The EPIC-pn data are produced using \texttt{epproc} and processed with the standard filtering criterion. We firstly extract 10--12~keV light curve using the \texttt{evselect} task. Time intervals with count rates exceeding 0.3 counts s$^{-1}$ are identified as periods affected by enhanced particle background and are excluded from the good time intervals (GTIs) used for the subsequent analysis. The EPIC-pn camera was operated in Timing mode. Source events are extracted from RAWX columns 29--46, centered on the source position. Background events are extracted from RAWX columns 3–7, which are located away from the source and free of source contamination. Only single and double events (\texttt{PATTERN<=4}) are used, and standard quality filtering (\texttt{FLAG==0}) is applied. The redistribution matrix file and ancillary response file are created by using the SAS tasks \texttt{rmfgen} and \texttt{arfgen}, respectively. 


\section{Results} \label{sec:results}

We analyze the spectrum using the X-ray spectral fitting package XSPEC 12.13.1 \citep{Arnaud1996}. We utilize the Wilm set of abundances \citep{Wilms2000} and Vern photoelectric cross sections \citep{Verner1996} in all fits. Cash statistics \citep{Cash1979} are employed to evaluate the fitting of RGS spectrum, while $\chi^2$ statistics are used for the EPIC-pn spectral fitting.


\begin{figure}
    \centering
    \includegraphics[width=1.0\linewidth]{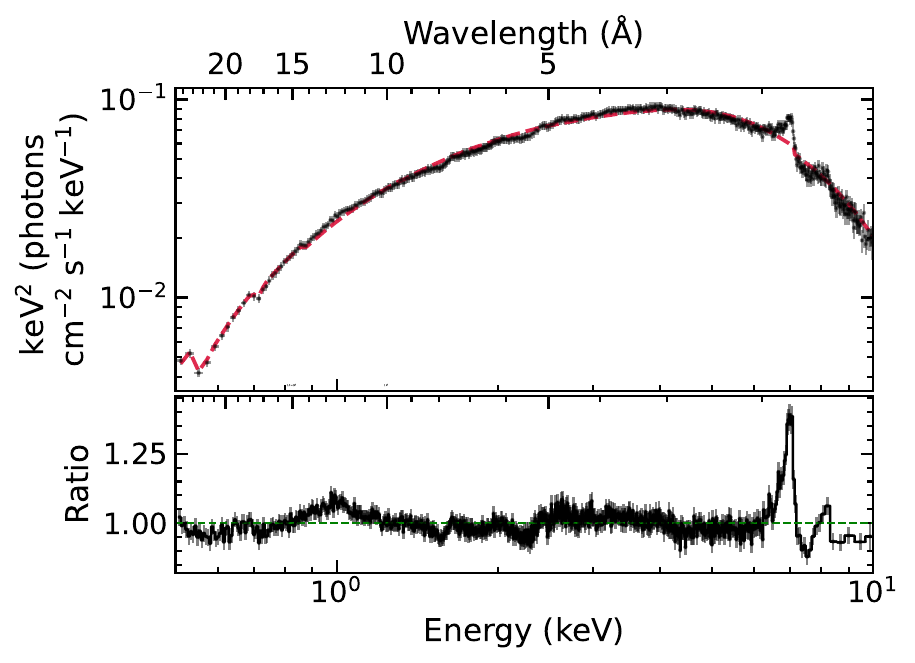} \\
    \caption{Top: EPIC-pn spectrum (black) with the best-fit continuum overlaid (red dashed). Bottom: data-to-model ratio of the fit with the continuum model \texttt{Tbabs*pcfabs*diskbb}. The 6.5--7.5~keV energy range was excluded when fitting the continuum model, but is included in the plotted spectrum and model for completeness.}
    \label{pn_spectrum}
\end{figure}


\subsection{Broadband spectrum and accretion state}

To investigate the accretion state during the observation, we extract the EPIC-pn spectrum to provide a broadband overview of the source’s X-ray emission. The continuum is well described by XSPEC model combination \texttt{Tbabs*pcfabs*diskbb}. Here, \texttt{TBabs} represents the photoelectric absorption in the interstellar medium, while \texttt{diskbb} corresponds to a multitemperature disk blackbody model \citep{Mitsuda1984, Makishima1986, Kubota1998}. The \texttt{pcfabs} component accounts for partial covering absorption toward the central engine, as suggested in some previous studies \citep{Shaw2022}.

To avoid potential biases introduced by the strong Fe emission features, the continuum fit is initially performed with the 6.5--7.5~keV energy range excluded. The continuum parameters reported below are obtained from this line-masked fit. Meanwhile, we verified that including this energy range in the fit does not significantly change the best-fitting continuum parameters, indicating that the inferred continuum properties are robust against the presence of the iron emission features.

The model combination provides a satisfactory description of the continuum emission, yielding a fit statistic of $\chi^2/\mathrm{dof} = 1930.87/1848$, as illustrated in Figure~\ref{pn_spectrum}. The fit indicates a high disk temperature of $kT_{\rm in} = 1.458\pm0.012$~keV (90\% confidence level), and an unabsorbed flux of $4.0\times10^{-10}$~ergs~cm$^{-2}$~s$^{-1}$ in the 13.6~eV--13.6~keV band. Assuming a source distance of 6.2~kpc and a mass of $6.4$~$M_{\bigodot}$ (\citealt{MacDonald2014, Orosz2001}), this corresponds to a luminosity of $1.8\times10^{36}$~ergs~s$^{-1}$, 0.3\% in terms of Eddington luminosity. This suggests that the source was in a soft state at the time of the observation, which is consistent with the report that the source remains in the soft state down to extremely low luminosities (e.g., \citealt{Hjellming2000, Connors2025}).

We also explored the inclusion of an additional Comptonization component using the \texttt{nthcomp} \citep{Zdziarski1996} model to account for possible coronal emission. However, the improvement in fit is marginal, with the value of $\chi^2$ reduced by only $<1$, indicating the weak contribution of the Comptonization component in the energy band of EPIC-pn. This is also consistent with the results of broadband NuSTAR observations \citep{Connors2025}. We therefore do not further pursue this component in the spectral modeling.

The best-fit parameters further suggest that only a small fraction of the disk is obscured, with the column density and the covering fraction in the \texttt{pcfabs} component of $\sim 3\times10^{23}$~cm$^{-2}$ and $\sim0.3$, respectively. It remains unclear whether the central engine in this system is heavily obscured. The extremely low X-ray flux observed in the soft state implies strong obscuration, but the broadband spectra do not support such a scenario (e.g. \citealt{Connors2025}). In the following discussion, we proceed under the assumption that no additional heavy obscuration is present.



\begin{figure*}
    \centering
    \includegraphics[width=0.9\linewidth]{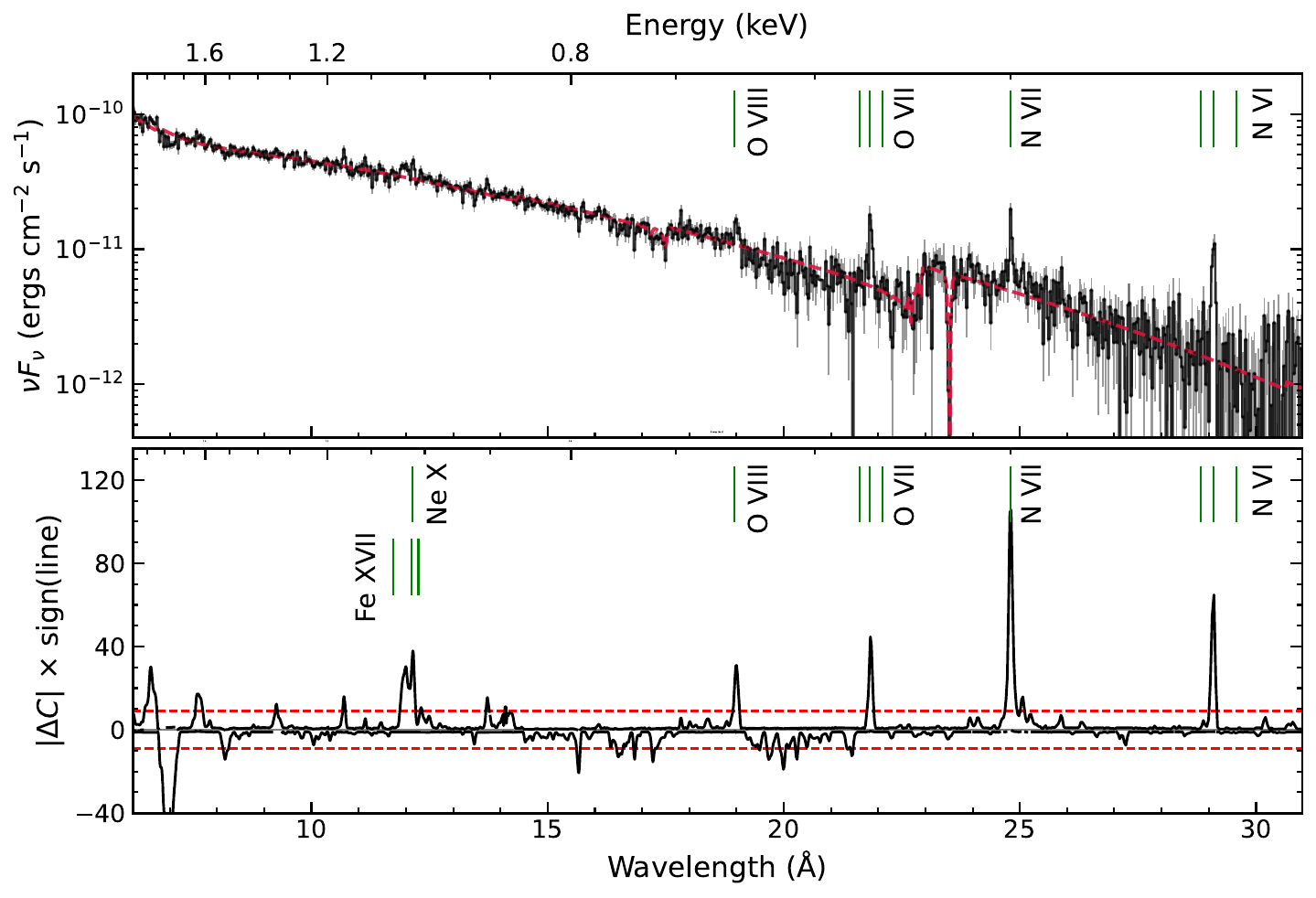} \\
    \caption{Upper panel: the RGS spectrum of \src\ in the 6--31~\AA\ range, shown in units of $\nu F_{\nu}$ (ergs cm$^{-2}$ s$^{-1}$). The data are shown in black with $1\sigma$ error bars, and the best-fit continuum model is overplotted as a red dashed line. Several strong emission lines are evident, including those from highly ionized N and O. The vertical green ticks mark the expected positions of known transitions, including O\,\textsc{viii} Ly$\alpha$, O\,\textsc{vii} He$\alpha$ triplet, N\,\textsc{vii} Ly$\alpha$, and N\,\textsc{vi} He$\alpha$ triplet. Lower panel: the result of a Gaussian line scan across the same wavelength range. The plotted quantity is the change in fit statistic ($\Delta C$) at each trial wavelength, multiplied by the sign of the Gaussian normalization. Horizontal dashed lines denote the $\pm 3 \sigma$ detection thresholds ($\Delta C = 9$). Several significant peaks are detected, consistent with known emission lines.}
    \label{pcf_dbb_residuals}
\end{figure*}

\subsection{Identification of emission lines}

Figure~\ref{pcf_dbb_residuals} presents the stacked RGS spectrum of the source. The underlying continuum is well described by a partially covered \texttt{diskbb} model, consistent with previous studies with EPIC-pn spectrum. Several prominent emission features are revealed. To characterize these spectral features, we first performed a Gaussian line scan to identify statistically significant lines, followed by full spectral modelling using a photoionization code.

To systematically search for emission features in the RGS spectrum, we perform a Gaussian line scan over the 6--31~\AA\ wavelength range using the method described by \cite{Psaradaki2018}. We add an unresolved Gaussian component (\texttt{gauss} in XSPEC) at each wavelength step to the smooth continuum model and record the corresponding improvement in the C-statistic. The Gaussian width is fixed at 0.005~\AA, and a scanning step size of 0.025~\AA\ is adopted. For this line scan, we reserve the continuum model of a partially covered disk blackbody, represented in XSPEC as \texttt{Tbabs*pcfabs*diskbb}. We also test an alternative model incorporating a partially covered power-law, but these produce negligible differences in the results. The result of this scanning procedure is shown in the bottom panel of Figure~\ref{pcf_dbb_residuals}, where the change in fit statistic ($\Delta C$) is plotted as a function of wavelength and the sign of the line normalization is preserved. Horizontal dashed lines indicate the $\pm3 \sigma$ detection thresholds ($\Delta C = 9$).


The scan reveals several statistically significant features. The strongest detections correspond to the known rest wavelengths of N\,\textsc{vii} Ly$\alpha$ at 24.78~\AA\ and the N\,\textsc{vi} He$\alpha$ at 29.08~\AA. Additional considerable features are observed at the expected locations of O\,\textsc{vii} He$\alpha$ at 21.80~\AA\ and O\,\textsc{viii} Ly$\alpha$ at 18.97~\AA. These lines are among the most prominent features typically observed in photoionized plasma (e.g. \citealt{Kallman2001}), and do not require a correction for the look-elsewhere effect. Line identifications and their reference wavelengths are made using AtomDB 3.1.3 \citep{Foster2012}.

Features with $\Delta C$ values above 9 are also observed at the locations around 12~\AA. This spectral region exhibits a complex blend of emission features, likely arising from multiple closely spaced transitions of ions such as Fe\,\textsc{xvii}, Ne\,\textsc{x}, and others. Due to the high density of lines in this band and the limited spectral resolution, it is challenging to identify individual transitions unambiguously. As such, in the following analysis, we do not focus on this region in detail and instead concentrate on the more precise and more isolated emission lines of N\,\textsc{vii}, N\,\textsc{vi}, O\,\textsc{viii}, and O\,\textsc{vii}. We note, however, that these features near 12~\AA\ are adequately reproduced in our photoionization modelling. At the same time, although several weaker peaks with $\Delta C$ around 9 are present, we do not treat these as statistically robust detections because of the large number of trial wavelengths tested, which would produce such features by chance (e.g. \citealt{Protassov2002}).


\subsection{Plasma diagnosing with emission lines}

To characterize the emission lines in a model-independent manner and to determine their kinematic properties, we perform a Gaussian line fitting analysis. We adopt a simple spectral model consisting of a partially covered disk blackbody component for the continuum and Gaussian components for the emission lines (i.e., N\,\textsc{vi}, N\,\textsc{vii}, O\,\textsc{vii}, and O\,\textsc{viii}). 

The O\,\textsc{viii} and N\,\textsc{vii} Ly$\alpha$ features are H-like Ly$\alpha$ transitions, which are fine-structure doublets consisting of the Ly$\alpha_{1}$ and Ly$\alpha_{2}$ lines. For example, in O\,\textsc{viii}, the two components are located at 18.967~\AA\ and 18.973~\AA, respectively. However, the wavelength separation is far below the resolving power of the RGS, and therefore the two components cannot be distinguished observationally. We therefore treat them as a single spectral feature and use the rest wavelength of the dominant Ly$\alpha_{1}$ component when deriving the velocity shifts listed in Table~\ref{tab:line_parameters}.

\begin{figure*}
    \centering
    \includegraphics[width=0.95\linewidth]{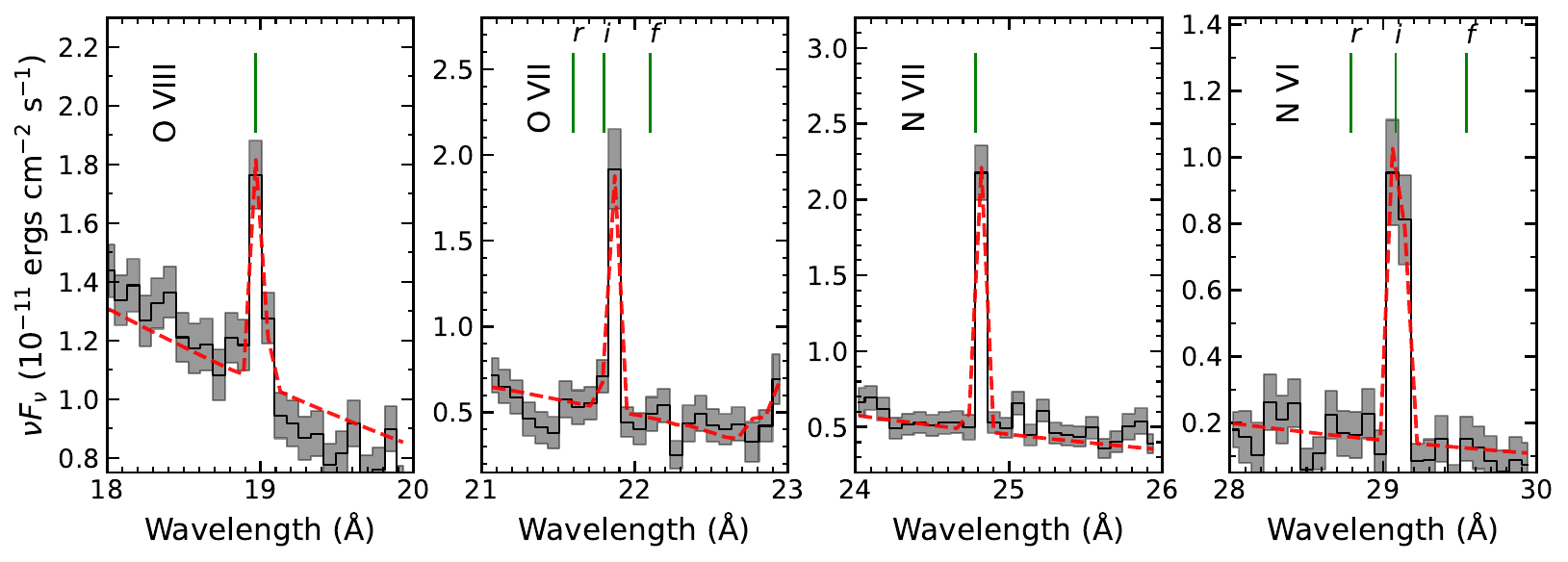} \\
    \caption{RGS spectrum of \src\ unfolded to best-fit model (Gaussian components plus the continuum). Each panel shows a region of the spectrum containing a likely line; the most probable transition, along with its rest wavelength, is indicated. The black histogram with grey error bars represents the observed spectrum, while the red dashed lines indicate the best-fit continuum plus Gaussian models.}
    \label{gauss_lines}
\end{figure*}

The emission features near 22~\AA\ and 29~\AA, corresponding to O\,\textsc{vii} and N\,\textsc{vi}, respectively, arise from the He-like triplets (e.g. \citealt{Porquet2000}). Each triplet is composed of a resonance ($r$) line, an intercombination ($i$) line, and a forbidden ($f$) line. For O\,\textsc{vii}, these lines are located at 21.602~\AA\ ($r$), 21.804~\AA\ ($i$), and 22.098~\AA\ ($f$); for N\,\textsc{vi}, they appear at 28.787~\AA\ ($r$), 29.084~\AA\ ($i$), and 29.535~\AA\ ($f$). Note that the intercombination feature is itself a fine-structure doublet, consisting of the so-called $x$ and $y$ lines, corresponding to the transitions ($1s2p\,^3P_2 \rightarrow 1s^2\,^1S_0$) and ($1s2p\,^3P_1 \rightarrow 1s^2\,^1S_0$), respectively \citep{Porquet2000}. For example, in O\,\textsc{vii}, the $x$ and $y$ lines are located at 21.801~\AA\ and 21.804~\AA, respectively. As in the case of the Ly$\alpha$ doublets, this wavelength separation is well below the RGS spectral resolution and the two components cannot be resolved. Consequently, they are not labeled separately in the figures. When calculating velocity shifts in Table~\ref{tab:line_parameters}, we adopt the wavelength of the dominant $y$ lines. Here, the intercombination ($i$) lines of both O\,\textsc{vii} and N\,\textsc{vi} are the strongest, as shown in Figure.~\ref{gauss_lines}. Since O\,\textsc{viii} and N\,\textsc{vii} are detected with, at most, a small wavelength shift, it is unlikely that the apparent intercombination lines are blue-shifted forbidden lines or redshifted resonance lines.

When fitting the emission lines with Gaussian components, we also include Gaussian components at the expected wavelengths of the resonance and forbidden lines to obtain upper limits on their line strengths for use in subsequent quantitative analysis. In these cases, the line widths and redshifts could not be well constrained, so we tie the widths and redshifts of the resonance ($r$), intercombination ($i$), and forbidden ($f$) lines to a common value. The corresponding line parameters, including line fluxes, widths, and velocity shifts, are summarized in Table~\ref{tab:line_parameters}.

The centroid wavelengths of the lines are set as free parameters to allow for possible bulk motion. By comparing the observed centroid wavelengths of the lines to their rest-frame values, the redshifts are determined. Separate redshift parameters are fitted for different lines. We find that all lines, including both nitrogen and oxygen species, exhibit consistent velocity shifts within uncertainties, indicating that the emitting plasma components share similar kinematics. 


From the measured fluxes of the N\,\textsc{vi} and O\,\textsc{vii} triplet lines, we compute the diagnostic line ratios G and R, which are commonly used to probe the physical conditions of emitting plasma \citep{Mewe1978, Liedahl1999, Porquet2000}. The G ratio, defined as $G = (f + i)/r$, is primarily sensitive to electron temperature, while the R ratio, defined as $R = f/i$, serves as a diagnostic for electron density. In these formulae, $f$, $i$, and $r$ represent the flux of the forbidden, intercombination, and resonance line, respectively. In our analysis, we performed plasma diagnostics using both the N\,\textsc{vi} and O\,\textsc{vii} He-like triplets. Since forbidden lines and resonance lines are not veritably detected for both N\,\textsc{vi} and O\,\textsc{vii}, the ratio R and G are only constrained as lower/upper limits. The G ratios derived from N\,\textsc{vi} and O\,\textsc{vii} are $>4.5$ and $>4.3$, respectively, both of which strongly indicate the photoionized nature of plasma \citep{Porquet2000, Porquet2010}. The corresponding R ratios are found to be $<0.14$ for N\,\textsc{vi} and $<0.35$ for O\,\textsc{vii}, implying an electron density of $>4 \times 10^{11}$~cm$^{-3}$, based on the theoretical predictions shown in Figure 8 of \cite{Porquet2000}. These diagnostics confirm that the emission arises in a dense, photoionized plasma.



\begin{table*}
\centering
\renewcommand\arraystretch{1.3}
\caption{Gaussian modelling parameters of the emission lines detected in the RGS spectrum. The columns give the line identification, normalization (in units of $10^{-5}$ ph cm$^{-2}$ s$^{-1}$), observed wavelength $\lambda_{\rm obs}$, rest-frame wavelength $\lambda_{\rm rest}$, Gaussian line width (\AA), the corresponding velocity dispersion $\sigma$, and the line-of-sight velocity shift $v_{\rm flow}$. Reported uncertainties correspond to 90\% confidence levels. Upper limits are given for undetected lines.}
\begin{tabular}{lccccccc}
\hline \hline
Line & norm ($10^{-5}$ ph cm$^{-2}$ s$^{-1}$) & $\lambda_{\rm obs}$ (\AA) & $\lambda_{\rm rest}$ (\AA) & Width (\AA) & $\sigma$ (km s$^{-1}$) & $v_{\rm flow}$ (km s$^{-1}$) \\
\hline
N\,\textsc{vii}         & $40 \pm 7$                      & $24.804^{+0.008}_{-0.009}$  & 24.779     &  $< 0.02$                    & $<227.3$                 & $310^{+90}_{-110}$ \\
N\,\textsc{vi}\textsubscript{r} & $<16$ &  -  & 28.787     & -               & -            & - \\
N\,\textsc{vi}\textsubscript{i} & $92^{+21}_{-20}$ & $29.097^{+0.014}_{-0.009}$        & 29.084     & $<0.03$ & $<309$ & $140^{+150}_{-90}$  \\
N\,\textsc{vi}\textsubscript{f} & $<10$ &  -  & 29.535     & -               & -            & - \\
\hline
O\,\textsc{viii}       & $12 \pm 4$                & $18.990^{+0.029}_{-0.012}$  & 18.967     & $< 0.06$       & $< 918$       & $360^{+500}_{-180}$ \\
O\,\textsc{vii}\textsubscript{r} & $<6$ & -        & 21.602     & -               & -            & - \\
O\,\textsc{vii}\textsubscript{i} & $35^{+10}_{-9}$                      & $21.842^{+0.010}_{-0.011}$         & 21.804      & $< 0.03$ & $< 343$ & $520^{+140}_{-150}$ \\
O\,\textsc{vii}\textsubscript{f} & $<9$ &  -       & 22.098     & -               & -            & - \\
\hline \hline
\end{tabular}
\label{tab:line_parameters}
\end{table*}

\subsection{Photoionized emission modelling}

In addition to constraining the plasma properties from individual lines, we fit the full spectrum with photoionization models to derive the physical properties. For this purpose, we employ the widely used photoionization code \texttt{XSTAR} (\citealt{Kallman2001}; \citealt{Kallman2004}), which calculates the ionization balance and level populations self-consistently under photoionization equilibrium. We note that the RGS spectrum shows only emission features, without clear absorption signatures. Given the intermediate to high inclination of \src\ \citep{Pahari2015}, the most straightforward and plausible interpretation is that the observed emission lines arise from scattering of the inner X-ray radiation by a disk atmosphere attached to the surface of the accretion disk. In this context, we adopt a slab geometry in our \texttt{XSTAR} calculations and use the reflected emission component for table model.

Instead of the RGS spectrum, we use the EPIC-pn data, which provide a broader energy coverage (0.5--10~keV), to constrain the ionizing continuum. The EPIC-pn continuum is well reproduced with a partially covered disk blackbody model. We assume the absorption-corrected best-fit continuum as the seed-ionizing spectrum. Simple Gaussian fits to the emission features yield a best-fit velocity broadening of approximately 150~km~s$^{-1}$, although the lower limit remains unconstrained (Table~\ref{tab:line_parameters}). This value is consistent with the results obtained from the \xrism\ observations of \src, where \citet{Parra2025} reported comparable line widths, especially for the emission lines at low-energy band (their Table~2), while likewise being unable to place a meaningful lower limit on the velocity broadening. Given the consistency between the \xmm\ and \xrism\ measurements, we adopt a turbulent velocity of 150~km~s$^{-1}$ for the \texttt{XSTAR} calculations to reduce the size of the tables. We fit four free parameters with \texttt{XSTAR} grid model, including number density ($n_{\rm H}$), column density ($N_{\rm H}$), ionization parameter ($\log \xi$), and redshift ($z$). 



\begin{figure*}
    \centering
    \includegraphics[width=0.95\linewidth]{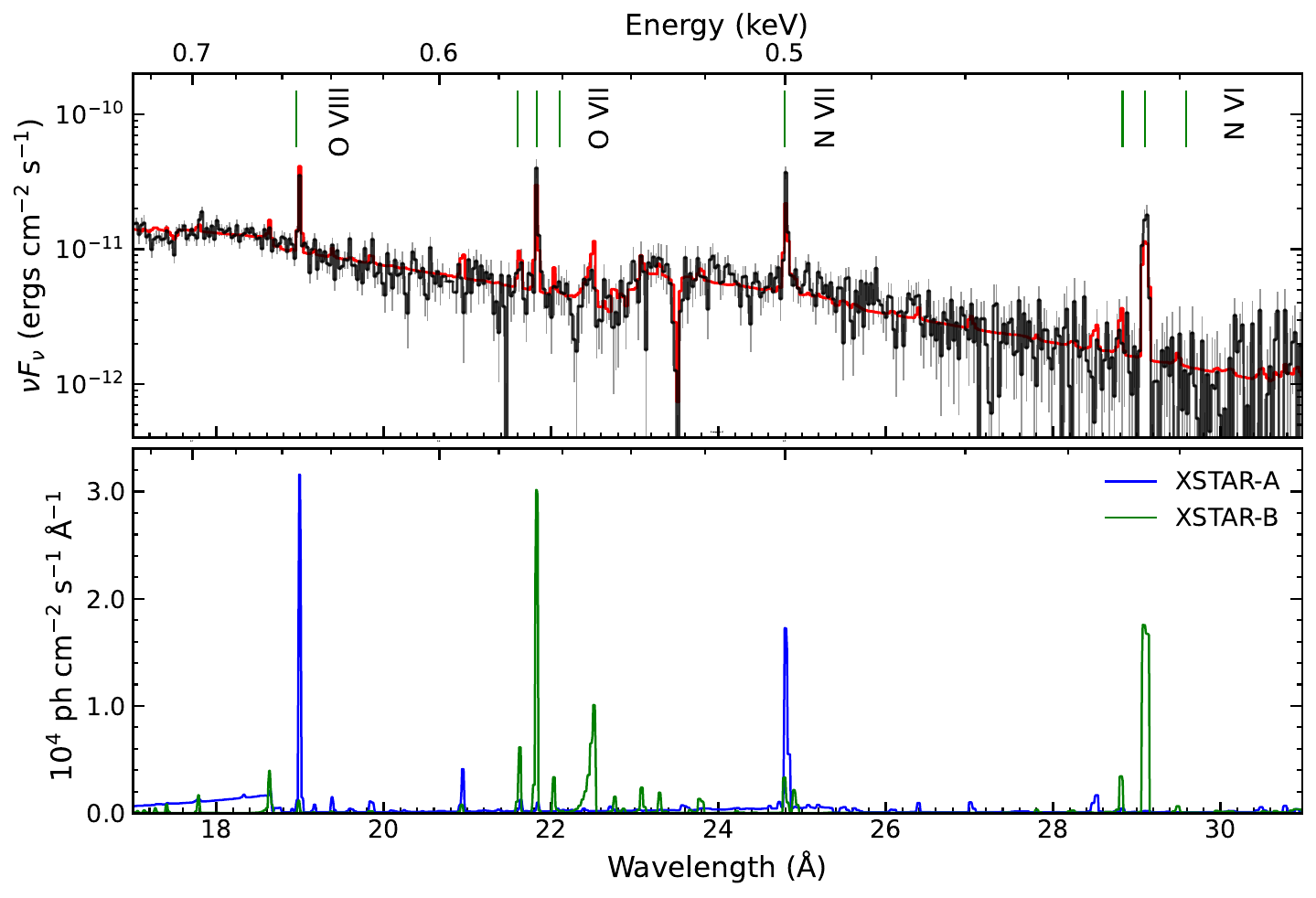} \\
    \caption{RGS spectrum and photoionized emission modelling. Top: observed spectrum (black) with the best-fit continuum+emission model overlaid (red). Vertical green ticks mark the rest wavelengths of prominent H- and He-like transitions. Bottom: spectra from two photoionized components (in photos cm$^{-2}$~s$^{-1}$~\AA$^{-1}$), illustrating the transitions that dominate two XSTAR components.}
    \label{rgs_spectrum}
\end{figure*}

A single photoionization component cannot simultaneously reproduce both intermediately ionized and less ionized lines (such as N,\textsc{vii} and N,\textsc{vi}). To account for this, we employ two \texttt{XSTAR} components with different ionization parameters to simultaneously reproduce them (hereafter referred to as component A and component B, respectively). A comparison between two \texttt{XSTAR} components is shown in the lower panel of Figure~\ref{rgs_spectrum}, while the best-fit model is shown in the upper panel of Figure~\ref{rgs_spectrum}. The corresponding best-fit parameters of the two photoionization components are listed in Table~\ref {xstar_table}. In Table~\ref {xstar_table}, we report the 90\% confidence intervals ($\sim 1.64 \sigma$) for all parameters, and additionally provide the 3$\sigma$ confidence ranges. This is done to minimize the potential contamination of weaker emission lines and the continuum of \texttt{XSTAR} components with respect to the robustness of the parameter constraints.


The two photoionized components, with ionization parameters ($\log \xi$) of 3.1 and 0.36, respectively, reproduce the emission lines of intermediately ionized nitrogen and oxygen ions as well as those of low ionized species. This implies that at least two distinct ionization zones are likely required to account for the observed spectrum. We measure redshifts of $\sim0.0024$ for component A and $\sim0.0018$ for component B, corresponding to plasma velocities of approximately $720$~km~s$^{-1}$ and $540$~km~s$^{-1}$, respectively. These measurements indicate a slowly flowing material directed away from the observer, aligning with the results with individual transition lines.

The N\,\textsc{vi} and O\,\textsc{vii} triplets associated with component B provide strong diagnostics of plasma density, as the ratio of the forbidden to intercombination line strengths is sensitive to electron density through collisional coupling between the upper levels \citep{Porquet2000}. In our modeling, we explore a density grid ranging from $10^{12}$ to $10^{16}$~cm$^{-3}$. The fitting yields a relatively high density of $2.2 \times 10^{14}$~cm$^{-3}$ for component B, consistent with the values inferred independently from the line-ratio analysis. In contrast, the density of component A remains poorly constrained, $<1.6 \times 10^{14}$~cm$^{-3}$ in $3\sigma$ confidence level, as the N\,\textsc{vii} and O\,\textsc{viii} lines are largely insensitive to density variations \citep{Kahn2002}.


\begin{table}
\centering
\renewcommand\arraystretch{1.3}
\caption{Best-fit parameters for two photoionization models.  Quoted uncertainties are intervals at 90\% confidence in the third column; the last column lists $3\sigma$ constraints.
\label{xstar_table}}
\begin{tabular}{lccc}
\hline\hline

Model  &  parameter  &  value(90\%)  &  value($3\sigma$)   \\ \hline
\texttt{XSTAR-A} & $n_{\rm H}$ [$10^{13}$~cm$^{-3}$] & $2.8_{-0.9}^{+3}$ & $<16$ \\
  & $N_{\rm H}$ [$10^{21}$~cm$^{-2}$] & $110_{-30}^{+50}$      & $110_{-60}^{+90}$     \\
  & $\log \xi$ & $3.1_{-0.4}^{+0.2}$ & $3.1_{-0.6}^{+0.4}$ \\
  & $z$ & $0.0024\pm0.0004$ & $0.0024\pm0.0007$ \\
  & Norm [$\times 10^{-5}$] & $6.3_{-1.0}^{+4}$ & $6_{-3}^{+12}$ \\  \hline
\texttt{XSTAR-B} & $n_{\rm H}$ [$10^{13}$~cm$^{-3}$] & $22_{-5}^{+5}$ & $22_{-9}^{+23}$ \\
  & $N_{\rm H}$ [$10^{21}$~cm$^{-2}$] & $3.9_{-2.2}^{+17.0}$   & $<58$ \\
  & $\log \xi$ & $0.361_{-0.12}^{+0.011}$ & $0.36_{-0.22}^{+0.03}$  \\
  & $z$ & $0.0018\pm0.0004$ & $0.0018\pm0.0007$ \\
  & Norm [$\times 10^{-5}$] & $27_{-7}^{+17}$ & $27_{-14}^{+40}$ \\   
\hline\hline
\end{tabular}
\end{table}


\section{Discussion} \label{sec:discussion}

In this work, we analyze the high resolution RGS spectrum of the black hole X-ray binary \src, and detect emission lines from N\,\textsc{vi}, N\,\textsc{vii}, O\,\textsc{vii}, and O\,\textsc{viii}. Using both Gaussian line fitting and photoionization modeling, we isolate the emission from the material surrounding the accretion disk. The results indicate the presence of a dense, photoionized medium, most likely associated with a hot disk atmosphere. We explore the implications of these findings for the accretion geometry and the outflow processes in \src\ in the following discussion.


\subsection{Physical properties of the emitting plasma}

The results from the Gaussian line fitting and the derived G-ratio strongly point to a photoionized nature of the emitting plasma. Furthermore, our full spectral modeling confirms that photoionization is the dominant mechanism of ionization. 



We find some evidence that the emission lines are redshifted, corresponding to a velocity shift of approximately 200--700~km~s$^{-1}$. However, this may partly arise from the absolute wavelength calibration of the RGS and from the radial velocity of the source. The absolute calibration uncertainty of RGS is about 6~m\AA, which corresponds to $\sim$700~km~s$^{-1}$ at the wavelength of N\,\textsc{vii} \citep{den_Herder2001}. This implies that the measured redshift may not reflect an intrinsic motion. In addition, the systemic radial velocity of the source could introduce a correction of 100--200~km~s$^{-1}$ \citep{Parra2025}. Allowing a conservative 200~km~s$^{-1}$ for radial motion of the source, our fits only will enable us to conclude that the emitter is effectively consistent with being static, or at most participating in inflow with velocities up to 900~km~s$^{-1}$ or even slow outflow. Note that when applying the photoionization model (\texttt{XSTAR}) to the full spectrum, the best-fit velocity shift slightly differs from that obtained with the Gaussian fits. Similar discrepancies between Gaussian fits to individual lines and full photoionization modeling have been reported in previous RGS and HETG studies (e.g., \citealt{Psaradaki2018}). In fact, this modeling bias becomes particularly evident when the velocity dispersion $\sigma$ is smaller than the instrumental resolution \citep{Cappellari2017}, as is the case in our study.

The fitting results in Table~\ref{xstar_table} indicate that the two photoionized components differ in ionization, density, and column density. The intermediate ionization component (A) is associated with a relatively low-density region ($n_{\rm H}\sim10^{13}$~cm$^{-3}$), whereas the low ionization component (B) originates in plasma with a density roughly one order of magnitude higher ($\sim10^{14}$~cm$^{-3}$). The distinction may point to a clumpy structure in the emitting plasma, but it is not statistically significant at the $3\sigma$ confidence level. 


Additional constraints on the geometry of the emitting region can be derived from the measured plasma parameters. Following \citet{Tarter1969}, the characteristic distance of the emitting plasma from the central source can be estimated as $R=\sqrt{L/n\xi}$, where $L$ is the ionizing luminosity, $n$ the electron density, and $\xi$ the ionization parameter. This relation holds under the assumption that there are no significant shielding effects present. Based on spectral fitting measurements, we get an estimation of the location of $\sim5\times10^{10}$~cm ($5\times10^{4}$~$R_{\rm g}$) and $\sim5\times10^{11}$~cm ($5\times10^{5}$~$R_{\rm g}$) for component A and component B, respectively. These values are broadly consistent with the characteristic scales inferred for the photoionized plasma components identified by \citet{Shaw2022}, who likewise found that the emitting gas is located far from the innermost accretion flow and is more naturally associated with an extended disk atmosphere or outflow.

The outer disk is expected to be tidally truncated at a radius of order $R_{\rm disk}\sim0.6-0.9$~$R_{\rm L,BH}$ \citep{Frank2002}, where $R_{\rm L,BH}$ is the radius of black-hole Roche lobe. Adopting published orbital parameters \citep{MacDonald2014}, we find that the accretion disk of \src\ would extend to $\sim5\times10^{11}$~cm ($5\times10^{5}$~$R_{\rm g}$). It indicates that the emitting plasma for component B is almost located at the outer edge of the accretion disk. These results also set a constraint on the location of the obscuring medium: if the central engine is indeed obscured, the medium must reside in the inner or middle regions of the accretion disk rather than in the outer disk or beyond.

Based on the disk-dominated spectrum suggested by EPIC-pn, the Compton radius is estimated to be $R_{\rm IC} \sim 6\times10^5 R_{\rm g}$ for a $6.4\,M_\odot$ black hole. The critical luminosity required to drive a thermal wind is estimated to be on the order of $L_{\rm crit} \sim 10^{37}\ \mathrm{erg\ s^{-1}}$. This estimate suggests that the emitting plasma we detect lies in region D of Figure~1 in \citet{Begelman1983}, corresponding to a quasi-static disk atmosphere. More accurate studies indicate that a thermal wind can be launched at radii of $\sim 0.2R_{\rm IC}$ (e.g., \citealt{Woods1996}). In this scenario, the emitting plasma associated with component B would fall within the wind-launching region. However, we do not detect any significant outflow velocity, and our current analysis does not provide a clear explanation for this discrepancy.

The intermediate ionization component (A) has a smaller plasma thickness ($N_{\rm H}/n_{\rm H}$), roughly three orders of magnitude smaller, compared to the low ionization component (B). This indicates the presence of two separate plasmas that cover different regions along the line of sight. Component A covers a broader area and extends closer to the central source, which naturally accounts for its higher ionization level. The low-ionization component B is probably vertically extended over the outer region of the disk, as suggested by a higher normalization. Such a structure is consistent with the layered geometry expected in accretion disk atmosphere, where the extended atmosphere is more extended in the outer part of the disk (e.g., \citealt{Psaradaki2018}). Although the RGS spectra do not allow us to constrain the line widths due to their limited spectral resolution, \xrism\ observation indicates that the emission lines are likely narrow, with widths of $\sim$200~km~s$^{-1}$ \citep{Parra2025}. These widths are even smaller than those expected from Keplerian broadening at the outer edge of the accretion disk. One possible scenario is that the emission originates in a failed disk wind region high up in the disk atmosphere, where the emitting material is located at larger cylindrical radii. In this case, the rotational velocities are reduced as a consequence of angular momentum conservation, leading to narrower line profiles (e.g. \citealt{Matthews2015}). Alternatively, the line profiles are modified by radiative transfer effects (e.g. \citealt{Munoz2026}).




\begin{figure}
    \centering
    \includegraphics[width=1.0\linewidth]{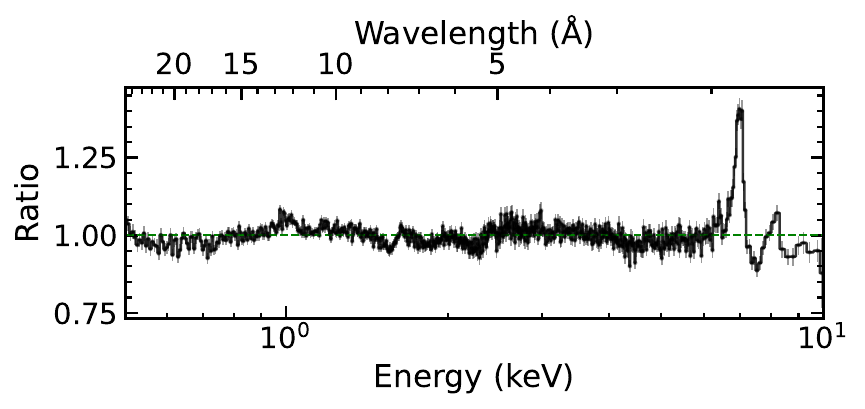} \\
    \caption{Residuals of EPIC-pn spectum to the best-fit RGS model with fixed parameters for the two \texttt{XSTAR} components. Strong excesses around 7.0~keV are clearly shown.}
    \label{pn_xstar_spectrum}
\end{figure}

\subsection{Additional photoionized component required?}

Observations with \chandra\ and \xrism\ (marked in Figure~\ref{maxi_lc}) have revealed the presence of several other highly ionized metal emission lines at shorter wavelength, including Fe\,\textsc{xxv} K$\alpha$ and Fe\,\textsc{xxvi} Ly$\alpha$ \citep{Shaw2022, Parra2025}. These components appear to participate in a complex inflow–outflow structure, the combination of significantly redshifted ($\sim700$~km~s$^{-1}$) and weakly blueshifted ($\sim-250$~km~s$^{-1}$) components, all varying strongly on ks timescales. The Analysis of \textit{NuSTAR} spectra further shows that these highly ionized emission lines (Fe\,\textsc{xxv} and Fe\,\textsc{xxvi}) are a persistent property of the source \citep{Connors2025}. 



We check the EPIC-pn spectrum obtained simultaneously with RGS to confirm potential highly ionized emission lines at shorter wavelengths. The emission-like features are significantly detected at $\sim7$~keV, as shown in Figure~\ref{pn_spectrum}, although the energy resolution of the EPIC-pn spectrum (especially in Timing mode) is insufficient to resolve the Fe\,\textsc{xxv} and Fe\,\textsc{xxvi} lines. We applied the same spectral model used for the RGS analysis to fit the EPIC-pn spectrum. The parameters of the two \texttt{XSTAR} components were fixed at the best-fitting values obtained from the RGS data, while the parameters describing the continuum were left free. The residuals to this model suggest a strong emission feature around 7.0~keV (Fe\,\textsc{xxvi}), as shown in Figure~\ref{pn_xstar_spectrum}. It suggests that an additional photoionization component with a higher ionization parameter is likely required to reproduce the emission features observed in the EPIC-pn spectrum. The higher ionization parameter is likely due to their extension to regions closer to the central source. The EPIC-pn instrument was operated in Window Timing mode during this observation. Since calibration uncertainties associated with Timing mode could bias detailed modeling, we present the EPIC-pn spectrum only as evidence for the presence of highly ionized iron emission lines, rather than performing full photoionization modeling.

\subsection{Comparison with previous \chandra\ observations}

Combining the RGS and EPIC-pn spectral analysis, the \xmm\ observation reveals three distinct photoionized plasma components characterized by low ($\log \xi \sim 0.36$), intermediate ($\log \xi < 3.1$), and high ionization states. A useful comparison can be made with the \chandra\ study of \citet{Shaw2022}, particularly their Epoch~2 observation, which was obtained at a comparable flux level. The two photoionized plasma components identified by \citet{Shaw2022} are likely associated with the intermediate and high ionization components detected in the present work.

Further support for this interpretation comes from the inferred plasma densities and emission radii. \citet{Shaw2022} derived a density of $\sim 5 \times 10^{12}\,\mathrm{cm^{-3}}$ for their first photoionized component, while our RGS analysis constrains the density of the intermediate ionization component to $<16 \times 10^{12}\,\mathrm{cm^{-3}}$ (lower limit $10^{12}\,\mathrm{cm^{-3}}$), consistent within the uncertainties. The inferred locations of the emitting plasma are also broadly compatible. For the intermediate ionization component, we derive an emission radius of $\sim 5 \times 10^{10}\,\mathrm{cm}$. \citet{Shaw2022} estimated a radius of $\sim 1.3 \times 10^{11}\,\mathrm{cm}$ under the assumption that the intrinsic luminosity exceeds the observed luminosity by a fiducial factor of 50 owing to obscuration. If the unobscured geometry assumed in the present work is adopted instead, the two estimates become comparable. 

The consistency between the RGS and \chandra\ constraints on the ionization state, density, and spatial location of the emitting plasma provides independent support for a accretion disk atmosphere or disk wind structure composed of multiple ionization zones. The \chandra\ analysis suggests that the highly ionized plasma is located closer to the central engine, whereas the RGS data indicate that the low ionization component originates at substantially larger radii. Combined with the intermediate ionization component detected in both datasets, these results point toward a radially stratified disk wind/atmosphere in which the ionization state decreases with increasing distance from the compact object. Notably, the low ionization component revealed by the RGS spectrum has no obvious counterpart in the \chandra/HETGS analysis, highlighting the importance of the longer wavelength coverage provided by \xmm/RGS.

\subsection{Radially extended disk atmosphere}

The properties of these emission lines, and the constraints on the physical conditions, indicate the presence of an accretion disk atmosphere located above the disk plane in \src. Soft X-ray emission and absorption features attributed to the disk atmosphere have been reported in several neutron star X-ray binaries, such as EXO 0748-676 \citep{Psaradaki2018} and Swift J1858.6-0814 \citep{Buisson2020}. Disk atmosphere signatures have also been suggested in the hard X-ray emission of several black hole X-ray binaries. For example, GRS 1915+105 shows quasi-static narrow absorption features consistent with disk atmosphere (\citealt{Lee2002}; \citealt{Ueda2009}). 


\cite{Diaz2016} and \cite{Munoz2026} presented summaries of low-mass X-ray binaries for which a photoionized plasma local to the source has been detected. Our results indicate that \src\ is a member of this growing population. The compilation of sources with disk atmosphere features shows that they are predominantly high-inclination systems (typically $i\ge65^\circ$). For instance, EXO~0748--676 \citep{Parmar1986}, 4U~1916--053 \citep{Trueba2020}, and MXB~1659--298 \citep{Iaria2018} are eclipsing or dipping neutron star binaries with inclinations above $70^\circ$, while black hole systems, including \src\ \citep{Pahari2015}, also show relatively high inclinations of $\sim65^\circ-80^\circ$. In such geometries, our line of sight intercepts or is located slightly above the outer layers of the accretion disk, the continuum being suppressed and making it easier to see emission lines. By contrast, in lower-inclination systems, the line of sight is more directly aligned with the central source and the inner disk, making it more difficult to detect these atmospheric features. This suggests that the presence of detectable disk atmosphere signatures in X-ray binaries is closely related to inclination, with high-inclination systems offering a privileged view of the disk’s upper layers and their photoionized plasma.


A notable distinction between our source and other black hole binaries is the detection of relatively low ionized emission lines at longer wavelengths. While such soft X-ray emission features have not been commonly reported in black hole systems \citep{Diaz2016}, similar low-ionization lines have been detected in neutron-star low-mass XRBs \citep{Psaradaki2018, Buisson2020}. This suggests that such components may not be unique to our source, but instead could be more widespread and underreported, potentially due to strong Galactic absorption that hampers their detection in many systems \citep{Buisson2020}. One possible explanation for the presence of these lines in our source is that it was observed in an extremely low-luminosity state, which allows low-ionization plasma to persist. In contrast, in more luminous systems, the intense radiation from the central engine is expected to maintain the disk atmosphere in a higher ionization state.

While the scenario of a disk atmosphere attached to the surface of the accretion disk provides a reasonable explanation for the observed emission features, we emphasize that the structure of the disk atmosphere and/or outflow is likely to be complex (e.g., \citealt{Miller2025}). In some black hole binaries, the disk atmosphere and disk winds are often observed to coexist (e.g., \citealt{King2015, Kallman2009}). In the RGS spectrum of \src, we do not detect any statistically significant blueshifted features, and therefore find no direct evidence for a fast disk wind. However, given the modest velocity resolution of the RGS, we cannot robustly distinguish between a quasi-static atmosphere and a low-velocity outflow. In particular, small velocity shifts of a few hundred km s$^{-1}$ would be difficult to detect with RGS and could remain hidden within systematic uncertainties. We also cannot entirely rule out the possibility that wind signatures are suppressed. For example, absorption features may become undetectable if the gas density is sufficiently high, or may be partially filled in by scattered emission from regions outside the direct line of sight.

Recent high-resolution observations with \xrism\ have revealed the presence of low-velocity X-ray winds in several systems, including cases where earlier data had been interpreted as arising from a static disk atmosphere (e.g., \citealt{Diaz2026}). As discussed in recent reviews \citep{Munoz2026}, some sources previously classified as having a disk atmosphere may instead host slow, weakly outflowing winds when observed with improved spectral resolution. In this context, while our results are consistent with a dense disk atmosphere, we cannot exclude the possibility that the absorbing material is part of a low-velocity wind.


Another possible alternative explanation is that these emission lines originate from photoionized stellar winds. This effect is most commonly observed in high-mass X-ray binaries, where intense X-ray emission from the compact object (neutron star or black hole) ionizes and heats the surrounding stellar winds from the companion star, producing strong H- and He-like emission lines as well as fluorescence features \citep{Stevens1990}. Vela X-1 is a prototype system hosting photoionized stellar winds \citep{Schulz2002, Watanabe2006, Diez2025}. However, \src\ does not fully match this scenario. Its companion is classified as a late B-type star (B9, and possibly as late as A0 within uncertainties), it is a giant rather than a supergiant. As a result, it is not expected to drive the strong stellar winds typically associated with B-type supergiants. Instead, the black hole accretion process appears to proceed via Roche-lobe overflow, classifying the system as a low-mass X-ray binary \citep{MacDonald2014}. In addition, both our Gaussian fitting and photoionization modeling indicate a high density plasma ($n_{\rm H} \sim 10^{13}$~cm$^{-3}$), whereas the expected stellar wind density of O/B-type stars at orbital scales is only $10^8-10^{10}$~cm$^{-3}$ \citep{Lamers1999, Kudritzki2000}. This discrepancy, together with the system configuration, argues against the interpretation of photoionized stellar winds in this case.


\subsection{Variable, multi-phase disk atmosphere structure}



The \xrism\ observations reported by \cite{Parra2025} were obtained at a substantially lower flux state than the \xmm\ observation analyzed in this work, with the X-ray flux approximately an order of magnitude lower. Under the simplifying assumption that the emitting gas is static, i.e. its density and distance from the ionizing source remain unchanged, the ionization parameter $\xi = L/(nr^2)$ \citep{Tarter1969} would scale linearly with luminosity. In that case, the plasma component observed here at $\log \xi \sim 3.1$ would be expected to shift to $\log \xi \sim 2.1$ during the \xrism\ observation. At such ionization levels, strong emission from ions such as S~\textsc{xvi}, S~\textsc{xv}  and Ar~\textsc{xviii} would be expected. However, no corresponding emission features (or only weak features) are detected in the \xrism\ spectrum. This absence suggests that the emitting outflow or disk atmosphere cannot be described as a static structure responding solely to changes in ionizing luminosity, but instead undergoes significant dynamical evolution.

We also note that \xmm\ observed \src\ during a previous soft-state episode, as shown in Figure~\ref{maxi_lc}. A detailed analysis of the RGS spectrum from that observation revealed no emission-line features similar to those discussed here. The EPIC-pn spectrum indicates that the X-ray flux during that observation was approximately two orders of magnitude higher than in the recent observation where the emission features were detected. If the disk atmosphere were static, the component observed in the current dataset at $\log \xi \sim 0.36$ would be expected to increase to $\log \xi \sim 2.36$, again assuming that $n$ and $r$ remain unchanged. At this ionization level, prominent emission from ions such as  O~\textsc{viii} and N~\textsc{vii} would be expected in the energy band of RGS. However, no such features are observed. The absence of these predicted emission lines further supports the conclusion that the disk atmosphere in \src\ is not static, but instead evolves dynamically in response to changes in the accretion flow.

\xrism\ observations of this source as well reveal that emission lines associated with outflows or disk atmosphere exhibit significant variability on timescales of a few kiloseconds \citep{Parra2025}. Motivated by this result, we search for similar behaviors in the RGS data. Derived for each 15~ks, time-resolved spectroscopy indicates a marginal change in the measured redshift of the emission lines between high- and low-flux intervals. However, we note that the velocity resolution of the RGS is approximately 700~km~s$^{-1}$ at this band. The observed differences in velocity are within this resolution limit, implying that they are not statistically significant. Consequently, we are unable to perform an analysis comparable to that enabled by \xrism\, and we do not pursue this further.

\section{Conclusions} \label{sec:conclusion}
In this work, we present an analysis of the most recent \xmm\ observation of the black hole X-ray binary \src. The high-resolution RGS spectrum reveals, for the first time, distinct soft X-ray emission lines that probably arise from the accretion disk atmosphere. We characterize the physical conditions of this atmosphere with both phenomenological Gaussian fitting and detailed photoionization modelling. The main findings of this study are summarized below:

\begin{enumerate}
\item High-significance detections of N\,\textsc{vi}, N\,\textsc{vii}, O\,\textsc{vii}, and O\,\textsc{viii} lines reveal a quasi-static or inflowing plasma.
\item He-like triplet diagnostics and photoionization modelling both indicate a dense, photoionized disk atmosphere with a density of $n_{\mathrm{e}} \sim 10^{12}-10^{15}$ cm$^{-3}$.
\item The RGS spectrum requires at least two photoionized components ($\log\xi \simeq 3.1$ and 0.36), while EPIC-pn Fe features imply an additional, highly ionized region.
\item The results point to a radially extended disk atmosphere viewed at a intermediate to high inclination ($\sim70^{\circ}$), consistent with similar findings in other X-ray binaries.
\item \src\ offers a unique opportunity to study the coupling between the disk atmosphere and outflows (X-ray and optical) at low luminosity.

\end{enumerate}

\section*{Acknowledgements}

The authors acknowledge and thank the \xmm\ team and Dr. Norbert Schartel for promptly observing the target through DDT. ZZ acknowledges support from ERC Synergy Grant 'Blackholistic' and the China Scholarship Council (CSC). RF acknowledges support from ERC Synergy Grant 'Blackholistic', The UKRI and The Hintze Family Charitable Foundation. JM acknowledges funding from a Royal Society URF. AA and TMD acknowledge support by the Spanish Agencia Estatal de Investigaci\'{o}n via PID2021-124879NB-I00 and PID2024-161863NB-I00. CB and JJ acknowledge support from the Warwick–Fudan Joint Seed Grant. MP acknowledges support from the JSPS Postdoctoral Fellowship for Research in Japan, grant number P24712, along with the JSPS Grants-in-Aid for Scientific Research-KAKENHI, grant number J24KF0244. This research has made use of MAXI data provided by RIKEN, JAXA and the MAXI team. 

\section*{Data Availability}

The \xmm\ data are available from the \xmm\ Science Archive (XSA) under the observation IDs listed in the paper. Calibrated products and analysis scripts will be shared upon reasonable request.



\bibliographystyle{mnras}
\bibliography{example} 





\bsp	
\label{lastpage}
\end{document}